\documentclass[12pt]{article}%
\usepackage{amsmath}
\usepackage{graphicx}
\usepackage{amsfonts}
\usepackage{amssymb}%
\providecommand{\U}[1]{\protect\rule{.1in}{.1in}}
\begin{document}

\title{The Gibbs Paradox and the Physical Criteria for the Indistinguishability of
Identical Particles}
\author{C. S. Unnikrishnan\\\textit{Gravitation Group, Tata Institute of Fundamental Research, }\\\textit{Homi Bhabha Road, Mumbai - 400 005, India}\\E-mail address: unni@tifr.res.in}
\date{}
\maketitle

\begin{abstract}
Gibbs paradox in the context of statistical mechanics addresses the issue of
additivity of entropy of mixing gases. The usual discussion attributes the
paradoxical situation to classical distinguishability of identical particles
and credits quantum theory for enabling indistinguishability of identical
particles to solve the problem. We argue that indistinguishability of
identical particles is already a feature in classical mechanics and this is
clearly brought out when the problem is treated in the language of information
and associated entropy. We pinpoint the physical criteria for
indistinguishability that is crucial for the treatment of the Gibbs' problem
and the consistency of its solution with conventional thermodynamics. Quantum
mechanics provides a quantitative criterion, not possible in the classical
picture, for the degree of indistinguishability in terms of visibility of
quantum interference, or overlap of the states as pointed out by von Neumann,
thereby endowing the entropy expression with mathematical continuity and
physical reasonableness.

\bigskip


\end{abstract}

\section{Multiple Notions of Entropy}

There is written record that J. von Neumann advised C. Shannon to call his
measure of information as entropy$^{1}$, saying, `in the first place, your
uncertainty function has been used in statistical mechanics under that name,
so it already has that name. In the second place, and more important, no one
knows what entropy really is, so in a debate you will always have the
advantage'. While the technical definitions consistent with both
thermodynamics, statistical mechanics and information theory are all known and
studied for decades, there is a remaining unease perhaps, exemplified by a
relatively recent and influential paper by E. T. Jaynes$^{2}$ on the Gibbs
paradox, in which he stated, `some important facts about thermodynamics have
not been understood by others to this day, nearly as well as Gibbs understood
them over 100 years ago...it is not surprising that entropy has been a matter
of unceasing confusion and controversy from the day Clausius discovered it.
Different people, looking at different aspects of it, continue to see
different things because there is still unfinished business in the fundamental
definition of entropy, in both the phenomenological and statistical
theories...further theoretical work will be needed before we can claim to
understand entropy.'

The high pace of quantum information theory has only complicated the scenario
further, with several new measures introduced and studied for quantities
associated with information encoded in quantum systems. However, in this paper
we focus on the limited problem called the Gibbs paradox, introduced by J. W.
Gibbs in the context of the change in entropy during mixing of gases. There
have been several discussions, and continued debates, on the notions of
entropy and the issue of indistinguishability of identical particles in the
context of the Gibbs paradox$^{3-6}$ (those cited here are indicative and far
short of being exhaustive). \ Central point of our discussion is the criterion
for indistinguishability of identical particles. We will argue that the
feature of indistinguishability is naturally present in classical physics,
without a need to impose it and then justify on the basis of the theory of
quantum mechanics. Most of our discussion will be within the conventional
understanding of entropy in thermodynamics and statistical mechanics, with
some reference to the simplest understanding of measure of information in the
sense of Shannon. We will also touch on the issue whether entropy is only a
theoretical entity, created by a definition as a useful and convenient notion
without a direct physical counterpart, or whether it has a well-constrained
relation to measurable physical quantities associated with the multi-particle system.

The notion of entropy in thermodynamics is associated with transfer of heat,
as introduced by Clausius, and the change in entropy in a thermodynamical
process from configuration $A$ to $B$ is defined through the expression%
\begin{equation}
\Delta S=\int_{A}^{B}\frac{dQ}{T}%
\end{equation}
where $dQ$ is the change in the quantity of heat. All real processes have some
generation of heat and the naturalness of entropy generation is what is
highlighted in the second law of thermodynamics. In the microscopic picture of
statistical mechanics of molecular processes, this then is well-understood as
distributing energy into available degrees of freedom at the molecular level
and this forms the basis of the Boltzmann definition of entropy%
\begin{equation}
S=k_{B}\ln W \label{Boltzmann}%
\end{equation}
where $W$ is the number of physical configurations of the multi-particle
system that is consistent with the given constraints on the macroscopic
variables like total energy. In this picture the entropy in a thermodynamic
state is calculated by counting the number of microscopic configurations that
corresponds to the macroscopic physical state.

The Gibbs' expression for entropy is identical (to factor $k_{B})$ in
structure with the information theoretic entropy,
\begin{equation}
S=-\sum_{i}p_{i}\ln p_{i} \label{Gibbs}%
\end{equation}
and this can be related to the Boltzmann entropy by noting that for equally
likely probabilities for the microstates accessible to the system in
equilibrium, $p_{i}=1/W_{i}.$

But, there is sometimes a gap between these microscopic notions and the
thermodynamic notion, which can lead to confusion or paradoxes. In fact, the
equivalence of the different definitions is theoretically demanded, and not
self-evident.$^{3}$ We note and stress here that only the Clausius definition
treats entropy as a physical quantity in terms of measurable entities of the
system and the other two definitions correspond to the theoretical
substructure that tries to understand the statistical distribution of physical
quantities in the multi-particle system. The connection of the Boltzmann and
Gibbs definitions to the physical entropy is through the frequentist picture
of probabilities and the number of configurations where the actual number of
particles with their physical attributes lying in specific range of values of
physical quantities can be related to the relevant probabilities.$^{7}$ Also,
an infinity of ensembles of the same physical system and ensemble averages are
theoretical concepts with no physical counterpart and the actual physical
system samples only a small fraction of accessible microstates in any finite
time, through its dynamics. However, the Gibbs definition allows
generalizations, the most discussed being the Shannon entropy in information
theory and its equivalents or extensions. The common feature in all notions of
entropy is the concept of sharing or distributing a physical resource among a
multi-system ensemble. \emph{The concept of sharing is primary and central,
and depending on which physical quantity is shared, the physical nature of
entropy varies}.

\section{The Gibbs Paradox and Gibbs' Solution}

The expression for entropy for  $n$ mole of gas, based on equation
\ref{Boltzmann}, could be $S=nR\ln V=Nk_{B}\ln V,$ up to a constant, and this
is of course expected to be additive (extensive) with change in volume and the
amount of gas, consistent with the expectation of the change in the number of
microscopic configurations available to the system. $N$ is the number of
particles. However, consider a situation where an imaginary partition is
inserted into a vessel containing the gas, dividing it equally into two. Now
the total entropy is
\begin{equation}
S=2\times\frac{N}{2}k_{B}\ln\frac{V}{2}=Nk_{B}\ln V-Nk_{B}\ln2
\end{equation}
Hence $\Delta S=-Nk_{B}\ln2,$ which is inconsistent with both thermodynamics
and physical expectation. No heat is transferred and no physical process took
place, except restricting access of the individual molecules to all available
microstates, by introducing a partition (locally, the molecules cannot  even
be `aware' of such a partition). The decrease in the entropy seems like the
violation of the second law of thermodynamics! The apparent discrepancy can be
made larger by further subdivisions. Instead if dividing by a partition, we
can also consider removing an already existing partition and then the entropy
increases by $Nk_{B}\ln2.$

On the other hand, if we consider the genuine mixing of two gases of different
species, each with $N$ molecules occupying volume $V/2$, by removing a
partition in a vessel, they mix spontaneously by expanding into volume $V$
(arranged to be at constant temperature) and the change in the entropy of
mixing is
\begin{equation}
\Delta S=2\times\left(  Nk_{B}\ln V-Nk_{B}\ln V/2\right)  =2Nk_{B}\ln2
\end{equation}
which is what one would expect because each molecule now has twice the number
of possibilities to occupy a compatible microstate, with increase in the
spatial volume available. However, this reasoning should not apply if the gas
is of the same species because then the partition can be considered as purely
imaginary, with no physical significance. There is nothing in the expression
for the entropy that can automatically take care of whether the molecules are
of the same species or not, or in other words, indistinguishable or not. This
is the origin of the Gibbs paradox.

Therefore, it is clear that the additional notion of `different species' or
distinguishability has to be introduced while calculating the entropy of
mixing of gases. This  was done by Gibbs and included in the counting of
microstates to solve the problem. However, the discussion on criteria for
indistinguishability, its justification, quantitative measures, subjectivity
etc. all occupied controversial discussions now well over a century. We may
now briefly recall Gibbs' solution to the problem.

The expression for the classical entropy is derived from $S=k_{B}\ln W,$ by
counting the number of distinguishable microstates for $N$ particles. The
number of microstates possible in a volume of phase space $\Delta
V_{ps}=\Delta p\Delta q$ is $W_{\Delta}=\Delta p\Delta q/h^{3},$ and indicates
that the entropy is proportional to the logarithm of the spatial volume.
Irrespective of the details of the counting to estimate $W$ in eq.
\ref{Boltzmann}, we note that if the counting of the number of states $W$ for
$N$ particles corresponds to an entropy $S=k_{B}\ln W=Nk_{B}\ln V,$ then
dividing this by an `indistinguishability factor' $N!$ changes the expression
for entropy to
\begin{align}
S &  =k_{B}\ln W^{\prime}=k_{B}\left(  N\ln V-N\ln N+N\right)  \nonumber\\
&  =Nk_{B}\ln(V/N)+k_{B}N
\end{align}
with Stirling's approximation $\ln N!\simeq N\ln N-N.$ With this expression we
can recalculate the difference between initial and final entropy when removing
(or introducing) a partition for the same species of gas, $N/2$ particles
occupying volume $V/2$ initially. We get,
\begin{align}
S_{i} &  =2\times\frac{N}{2}k_{B}\ln(\frac{V/2}{N/2})+2\times k_{B}%
(N/2)\nonumber\\
S_{f} &  =Nk_{B}\ln(V/N)+k_{B}N=S_{i}%
\end{align}
Therefore, the Gibbs' division by the factor $N!$ on the number of
microstates, invoking indistinguishability of identical particles, eliminates
the unphysical increase of entropy for the same species of gas. For the case
of two different species of gas mixing we get
\begin{align}
S_{i} &  =2\times\frac{N}{2}k_{B}\ln(\frac{V/2}{N/2})+2\times k_{B}%
(N/2)\nonumber\\
S_{f} &  =(N/2)_{1}k_{B}\ln V/(N/2)_{1}+(N/2)_{2}k_{B}\ln V/(N/2)_{2}%
+k_{B}(N/2)_{1}+k_{B}(N/2)_{2}\\
\Delta S &  =S_{f}-S_{i}=Nk_{B}\ln2
\end{align}
where we have labelled the two different gases with subscripts in the
expression for $S_{f}$. The result agrees with the physical expectation for
the entropy of mixing of different gases when a molecule of a species has
twice as many microstates accessible. Thus, the Gibbs' division by $N!$ seems
to be correct solution. It only remains to justify the division on physical
grounds of indistinguishability. Perhaps the main reason for the original
counting of states with particles treated as distinguishable is that every
configuration of the particles is notionally distinguishable by the Newtonian
history, due the existence of trajectories and that particle exchanges can be
kept track of. Therefore, most people think that indistinguishability of
identical particles is justified only within quantum theory and that quantum
theory as a basis is essential for the correct treatment of statistical
mechanics even at the level of non-degenerate gases, at low density where
interparticle separation is much larger than the de Broglie wavelength. If
this were true, it would be surprising -- this anticipation of quantum theory
in a simple thermodynamical problem. It is to this aspect we now turn the
attention for a careful analysis.

\section{Counting Microstates for Entropy}

Given $N$ \ distinguishable particles, the number of microstates with $n$
particles in a partition is%
\begin{equation}
W(N,n)=\frac{N!}{n!(N-n)!}%
\end{equation}
If there is additional degeneracy of $g$ (number of spin states, for example),
this is multiplied by the factor $g^{n}.$ $N$ particles can be distributed in
allowed physical states (of energy, spin projection etc.) labelled $1,2...m$
with partitions $n_{1},$ $n_{2}...n_{m}$ in number of ways
\begin{equation}
W(N,n_{i})=\frac{N!g_{1}^{n_{1}}g_{2}^{n_{2}}..g_{m}^{n_{m}}}{n_{1}%
!n_{2}!...n_{m}!}=\frac{N!\prod\nolimits_{i}g_{i}^{n_{i}}}{\prod n_{i}!}
\label{ways}%
\end{equation}
With the Stirling approximation,
\begin{equation}
S/k_{B}=\ln W=N\ln N+\sum_{i}n_{i}\ln(g_{i}/n_{i})
\end{equation}
The probability for a particle to be in the partition $i$ is $n_{i}%
/N=g_{i}\exp(-\epsilon_{i}/k_{B}T)/Z,$ where the partition function $Z=\sum
g_{i}\exp(-\epsilon_{i}/k_{B}T).$ Therefore,
\begin{equation}
\frac{g_{i}}{n_{i}}=\frac{Z}{N}\exp\epsilon_{i}/kT
\end{equation}
from which%
\begin{equation}
S/k_{B}=\ln W=N\ln Z+U/k_{B}T
\end{equation}
Since $Z\propto VT^{3/2}$ for the gas, the entropy is
\begin{equation}
S=Nk_{B}\ln V+\frac{3}{2}kT+C
\end{equation}
(with $C$ a constant) and it is this expression that leads to the Gibbs
paradox. Dividing the expression for $W(N,n_{i})$ by $N!$ gives%
\begin{align}
S/k_{B}  &  =N\ln V+\frac{3}{2}kT-N\ln N+N+C\nonumber\\
&  =N\ln(V/N)+\frac{3}{2}kT+N+C
\end{align}
When the partition is inserted in the vessel containing a gas, we have%
\begin{align}
S_{i}  &  =N\ln\left(  V/N\right)  +C^{\prime}\nonumber\\
S_{f}  &  =2\times\frac{N}{2}\ln\left(  \frac{V/2}{N/2}\right)  +C^{\prime
}\nonumber\\
\Delta S  &  =0
\end{align}
avoiding the Gibbs paradox.

The counting corresponding to the indistinguishable particles could have been
done directly, from the outset, by noting that the number of possibilities of
distribution of $n_{i}$ particles in g-fold degenerate state with
$g_{i}>>n_{i}>>1$, as often is the case in the thermodynamic problem, is%
\begin{align}
W_{i}  &  =\frac{\left(  n_{i}+g_{i}-1\right)  !}{n_{i}!(g_{i}-1)!}\\
&  =\frac{\left(  n_{i}+g_{i}-1\right)  \left(  n_{i}+g_{i}-2\right)
...\left(  n_{i}+g_{i}-n_{i}\right)  _{n~terms}\times(g_{i}-1)(g_{i}%
-2)...1}{n_{i}!(g_{i}-1)!}\\
&  \simeq\frac{g_{i}^{n_{i}}(g_{i}-1)!}{n_{i}!(g_{i}-1)!}%
\end{align}
and
\begin{equation}
W=\prod\nolimits_{i}W_{i}\simeq\prod\nolimits_{i}\frac{g_{i}^{n_{i}}}{n_{i}!}%
\end{equation}
without the factor $N!,$ when compared to equation \ref{ways}.

Now we turn to the crucial issue of understanding and quantifying the notion
of indistinguishability.

\section{Indistinguishability: Physical Aspects}

In the thermodynamical problem of a gas in a container at moderate densities
and at temperatures well above the absolute zero temperature, the quantum
mechanical aspects of indistinguishability can play no role. It is only when
the de Broglie wavelength is comparable to the interparticle separation, one
expects quantum mechanical considerations to be important. The manifestation
of indistinguishability in quantum physics is interference of amplitudes and
in problems that does not have to deal with interference and quantum
correlations, quantum theoretical aspects are irrelevant. Therefore, it is
physically unreasonable to invoke quantum mechanical justification for
treating particles as indistinguishable in the Gibbs entropy problem. This
means that it will be truly an advance in conceptual understanding if we could
justify why we need to treat classical gas as consisting of indistinguishable particles.

Another aspect that is important in the solution of the Gibbs problem is that
the usual discussions of indistinguishability smells of some subjectivity in
determining whether a particle is distinguishable from another. What decides,
and by what criteria, whether particles are distinguishable? Nature should not
care whether the physicist is able to distinguish one particle from another,
with his tools and methods. The notion of physical entropy change should be
free of such subjectivity, unless entropy is merely a theoretical construct
meant for determining the amount of useful work, based on available subjective
knowledge about the microscopic details of the system. Jaynes' paper discusses
this point in great detail, in support of such subjectivity$^{2}$. The problem
is only amplified when one deals with physical indistinguishability in the
language of information, because the notion of information is associated with
acquisition and representation of information, which has subjective aspects
built in.

As we already remarked, Newtonian trajectories of classical particles is one
reason such particles are treated as tractable and therefore distinguishable.
However, for a specifying a particular equilibrium physical configuration in
phase space (with particular coordinates and momenta at any instant for $N$
particles) the history is not relevant at all. Even if we can keep track of
the trajectories and histories of particles and even label individual
particles (in a computer program, for example), the configurations with only
permutations of particles are \emph{not physically distinct even though
symbolically distinct}. The crucial point is that if a physical interaction
cannot distinguish one particle from another, in principle, they are
indistinguishable. Distinguishability is equivalent to spatial separability
and this single criterion is sufficient to show that indistinguishability of
identical particles is already a classical feature. This has really nothing to
do with quantum mechanics of the particles, except in the general sense of
quantum theory being the general theory of description of particles and their
interactions. There is no subjectivity either, because what matters is not
whether the analytical physicist has enough information available about the
particles to decide whether they are distinguishable or not, but whether they
behave differently under a general set of physical fields acting on them. This
criterion focuses on the intrinsic properties of the particle, and not on
whether or not one has information about such properties.

It is easy to see that classical theory of representation of information with
classical symbols do not treat multiple occurrences of the same symbols as
distinguishable. For $n$ classical bits with each bit distributed among $g$
states (equivalent to the degeneracy in the particle case), the number of
classical states is $W_{cl}=g^{n}$ and not $n!g^{n}.$ Therefore, trying to
enumerate the accessible number of configurations for $N$ particles with
$n_{i}$ particles distributed in state with degeneracy $g_{i}$ as
$W=N!\prod\nolimits_{i}g_{i}^{n_{i}}/n_{i}!$ was flawed from the beginning,
and there is no need to invoke quantum mechanics to justify the
indistinguishability. Instead of division by $N!,$ what was really required
was the correct way of enumeration of physically distinct states in the first
place, instead of mentally distinct states.

\section{Indistinguishability in Quantum Mechanics}

The need for quantum theoretical analysis enters considerations of statistical
mechanics and thermodynamics only when the particles have to compete for
available phase space, when quantum degeneracy becomes important and manifest
in the interference of relevant amplitudes. In that regime, exchange of
particles' coordinates have physical consequences, like spin-statistics
correlations. While indistinguishability and exchange of identical particles
have physical consequences in a quantum physical context, it has no observable
physical consequence in classical physics. \ However, the physical feature of
indistinguishability of identical particles is a basic notion that is
independent of the theoretical framework, just as the indistinguishability of
identical alphabets of languages or mathematical symbols. (It is not really
frivolous or naive to say that when such indistinguishable classical symbols
themselves are constructed out of molecular systems, one has no justification
to treat the molecules themselves as distinguishable!) Indistinguishability
signifies the impossibility of selective segregation and filtering by any
physical means whatsoever$^{2,3}$ (this connects the Gibbs problem to the
Maxwell's demon problem).

Now we address the important issue of quantifying the degree of
distinguishability. This has already been clarified fully by von Neumann$^{8}%
$, while discussing the (related) problem of the Maxwell's demon. In the
discussion of the Gibbs problem of entropy of mixing of two gases with equal
number and volume, it seems that even the slightest of distinguishability
implies the change of entropy amounting to $Nk_{B}\ln2,$ which is not
satisfactory from a physical point of view. It is in fact this feature that
lead to the discussion on subjectivity of change of entropy. However, as von
Neumann pointed out, two physical states are truly distinguishable in physics
only when they are orthogonal in the quantum mechanical sense. If the physical
states are represented as $\left\vert a\right\rangle $ and $\left\vert
b\right\rangle ,$ the degree of overlap and indistinguishability, is
$\left\vert \left\langle a|b\right\rangle \right\vert $ and there is no
corresponding notion in classical physics. In classical physics,
distinguishability is a discontinuous concept, whereas in quantum physics it
is continuous. This is important in treating the problem of entropy of mixing
without the \ discontinuity apparent in the classical discussion. It is
perhaps best to quote directly von Neumann$^{8}$ (with slight change of notation),

`In particular, the above treatment shows that two \ states $\left\vert
a\right\rangle $ and $\left\vert b\right\rangle $of the system S can be
certainly divided by \ a semi-permeable wall if they are orthogonal. We now
want \ to prove the converse: if $\left\vert a\right\rangle $ and $\left\vert
b\right\rangle $ are not orthogonal, then \ the assumption of such a
semi-permeable wall contradicts \ the second law of thermodynamics. That is,
the necessary \ and sufficient condition for the separability by
semi-permeable walls is $\left\langle a|b\right\rangle =0$ , and not, as in
classical \ theory, $\left\vert a\right\rangle \neq\left\vert b\right\rangle
.$ This clarifies an old paradox of the classical form of thermodynamics,
namely, the uncomfortable \ discontinuity in the operations with
semi-permeable walls: \ states whose differences are arbitrarily small are
always \ 100\% separable, the absolutely equal states are in general \ not
separable! We now have a continuous transition: It \ will be seen that 100\%
separability exists only for $\left\langle a|b\right\rangle =0$ and for
increasing $\left\langle a|b\right\rangle $ it becomes steadily worse.
\ Finally, at maximum $\left\langle a|b\right\rangle ,$ i.e., $\left\langle
a|b\right\rangle =1$, the \ states $\left\vert a\right\rangle $ and
$\left\vert b\right\rangle $ are identical, and the separation is completely impossible.'

These considerations imply that the expression for the change in entropy of
mixing, $\Delta S=Nk_{B}\ln2$ in the case of two equal partitions considered
earlier, should be multiplied by the visibility function $\left\vert
\left\langle a|b\right\rangle \right\vert ^{2}$ to completely resolve the
Gibbs paradox. The change in entropy does depend on the degree of similarity
or the degree of indistinguishability and it can be precisely formulated. The
von Neumann approach also can take care of correlations in the system and
provides a truly general setting to deal with both classical and quantum
entropy, and hence information theoretic entropy as well.

\section{Connecting to Clausius Entropy}

The clarifications on the entropy of mixing and separation in terms of true
physical distinguishability allows us to relate the discussion to the Clausius
definition of change of entropy in a thermodynamical process, $\delta
S=\int\delta Q/T,$ where $Q=U-F,$ where $U$ is the internal energy and $F$ the
free energy. The question of `how hard' is it to separate the already mixed
gas is the central feature that encodes the salient point of our discussions.
In the classical treatment, the (discontinuous) transition from
indistinguishable to distinguishable is equivalent to the possibility of
restoring the mixed gases back into the partitions by applying an external
field, without the physicists having to play the demon, which in any case is
impossible as shown in several discussions of Maxwell's demon. This is the
natural and physical definition of distinguishability. This transition is made
smooth and quantitative in the quantum theoretical approach. The motions that
result from the external field leads to the separation and this in turn
changes the external fields. \emph{Thus, distinguishable micro-physical states
are equivalent to the possibility of attaining distinguishable macroscopic
states}. This is the key point and it can be related to the change in the
quantity of heat and the change in physical entropy.

As an example, consider spin-1/2 atoms of the same species, at low density and
finite temperature, in a container with a partition. In the absence of an
external magnetic field the fact that there are two distinguishable
projections for the spin on a reference axis has no physical relevance and the
mixing of the gas does not result in a change of the macroscopic configuration
or physical quantities. Nor does it alter the phase space available to each
particle. Hence, the entropy should not change in the absence of a field, in
spite of the notional distinguishability. However, if the two spin states were
separated into two partitions to start with, in the presence of a field,
mixing should result in a change in the physical and mathematical entropy,
from the considerations we discussed in this paper.

As a concluding remark, it seems that we need to only consolidate the various
insights discussed on the concepts of entropy through ages, from Clausius to
Shannon, and there doesn't seem to be compelling reasons to think that `there
is still unfinished business in the fundamental definition of entropy, in both
the phenomenological and statistical theories' as Jaynes remarked. His remarks
were in the context of considering entropy as a theoretical construct,
necessary to evaluate the amount of useful work available from the
thermodynamical system, which is subjective in the sense that more information
about the microstates of the system leads to a better of estimate of entropy
and practical strategy to extract useful work. However, entropy as the unique
physical quantity that attains the maximum value in the multi-particle
thermodynamic system at its equilibrium does not depend on such subjective
knowledge and information. There is no true correspondence between information
that is practically available about the system and its physical entropy. While
there is scope to define and use new quantities that resemble physical entropy
in different contexts involving statistical considerations, the concept and
definition of thermodynamical and statistical mechanical entropy, endowed with
properties like extensivity and continuity due to insights from Gibbs and von
Neumann, seems to be on a robust and consistent foundation.

\section*{Acknowledgments}

This paper evolved from talks at the School of Language, Logic and Reality at
the IIT, Jodhpur and at the International Program on Quantum Information at
the Institute of Physics, Bhubaneswar, now a regular platform for discussions
on foundational aspects of quantum theory. I thank Subhashish Banerjee and
Debajyoti Gangopadhyay, and Pankaj Agarwal and Arun Pati, the friendly
organizers of the two meetings, for the opportunity and help.

\bigskip

\bigskip

\bigskip

\bigskip
\end{document}